\documentclass[preprint,prd,aps,showpacs,showkeys,nofootinbib]{revtex4}
\usepackage{graphicx}
\begin{document}


\title{Higgs boson mass corrections in the $\mu\nu$SSM with effective potential methods}

\author{Hai-Bin Zhang$^a$\footnote{email:hbzhang@hbu.edu.cn},
Tai-Fu Feng$^a$\footnote{email:fengtf@hbu.edu.cn},
Xiu-Yi Yang$^b$,
Shu-Min Zhao$^a$,
Guo-Zhu Ning$^a$}

\affiliation{$^a$Department of Physics, Hebei University, Baoding, 071002, China\\
$^b$College of Science, University of Science and Technology Liaoning, Anshan, 114051, China
}

\begin{abstract}
To solve the $\mu$ problem of the MSSM, the $\mu$ from $\nu$ Supersymmetric Standard Model ($\mu\nu$SSM) introduces three singlet right-handed neutrino superfields $\hat{\nu}_i^c$, which lead to the mixing of the neutral components of the Higgs doublets with the sneutrinos, producing a  relatively large CP-even neutral scalar mass matrix. In this work, we analytically diagonalize the CP-even neutral scalar mass matrix and analyze in detail how the mixing impacts the lightest Higgs boson mass. We also give an approximate expression for the lightest Higgs boson mass. Simultaneously, we consider the radiative corrections to the Higgs boson masses with effective potential methods.
\end{abstract}


\maketitle

\section{Introduction\label{sec1}}
\indent
Since the ATLAS and CMS Collaborations reported the significant discovery of a new neutral Higgs boson~\cite{ATLAS,CMS}, the Higgs boson mass is now precisely measured by~\cite{ATLAS-CMS}
\begin{eqnarray}
m_h=125.09\pm 0.24\: {\rm{GeV}}.
\end{eqnarray}
Therefore, the accurate Higgs boson mass will give most stringent constraints on parameter space for the standard model and its various extensions.

As a supersymmetric model, the ``$\mu$ from $\nu$ supersymmetric standard model'' ($\mu\nu$SSM) has the superpotential:~\cite{mnSSM,mnSSM1,mnSSM2,ref-zhang1,ref-zhang-LFV,ref-zhang2,ref-zhang-HLFV}
\begin{eqnarray}
&&W={\epsilon _{ab}}\left( {Y_{{u_{ij}}}}\hat H_u^b\hat Q_i^a\hat u_j^c + {Y_{{d_{ij}}}}\hat H_d^a\hat Q_i^b\hat d_j^c
+ {Y_{{e_{ij}}}}\hat H_d^a\hat L_i^b\hat e_j^c + {Y_{{\nu _{ij}}}}\hat H_u^b\hat L_i^a\hat \nu _j^c \right)
\nonumber\\
&&\hspace{0.95cm}-  {\epsilon _{ab}}{\lambda _i}\hat \nu _i^c\hat H_d^a\hat H_u^b + \frac{1}{3}{\kappa _{ijk}}\hat \nu _i^c\hat \nu _j^c\hat \nu _k^c \:,
\label{eq-W}
\end{eqnarray}
where $\hat H_u^T = \Big( {\hat H_u^ + ,\hat H_u^0} \Big)$, $\hat H_d^T = \Big( {\hat H_d^0,\hat H_d^ - } \Big)$, $\hat Q_i^T = \Big( {{{\hat u}_i},{{\hat d}_i}} \Big)$, $\hat L_i^T = \Big( {{{\hat \nu}_i},{{\hat e}_i}} \Big)$ are $SU(2)$ doublet superfields, and $Y_{u,d,e,\nu}$, $\lambda$, and $\kappa$ are dimensionless matrices, a vector, and a totally symmetric tensor, respectively.  $a,b=1,2$ are SU(2) indices with antisymmetric tensor $\epsilon_{12}=1$, and $i,j,k=1,2,3$ are generation indices. The summation convention is implied on repeated indices in this paper.
Besides the superfields of the MSSM~\cite{MSSM,MSSM1,MSSM2,MSSM3,MSSM4}, the $\mu\nu$SSM introduces three singlet right-handed neutrino superfields $\hat{\nu}_i^c$ to solve the $\mu$ problem~\cite{m-problem} of the MSSM. Once the electroweak symmetry is broken (EWSB), the effective $\mu$ term $-\epsilon _{ab} \mu \hat H_d^a\hat H_u^b$ is generated spontaneously through right-handed sneutrino vacuum expectation values (VEVs), $\mu  = {\lambda _i}\left\langle {\tilde \nu _i^c} \right\rangle$. Additionally, three tiny neutrino masses can be generated at the tree level through a TeV scale seesaw mechanism~\cite{mnSSM,mnSSM1,mnSSM2,ref-zhang1,ref-zhang-LFV,ref-zhang2,meu-m,meu-m1,meu-m2,meu-m3,
neu-zhang1,neu-zhang2,ref-zhang3}.

In the $\mu\nu$SSM, the left- and right-handed sneutrino VEVs lead to the mixing of the neutral components of the Higgs doublets with the sneutrinos producing an $8\times8$ CP-even neutral scalar mass matrix, which can be seen in Refs.~\cite{mnSSM1,mnSSM2,ref-zhang1}. Therefore, the mixing would affect the lightest Higgs boson mass. In this work, we analytically diagonalize the CP-even neutral scalar mass matrix, which would be conducive to the follow-up study on the Higgs sector. In the meantime, we consider the Higgs boson mass corrections with effective potential methods. We also give an approximate expression for the lightest Higgs boson mass. In numerical analysis, we will analyze how the mixing affects the lightest Higgs boson mass.

Our presentation is organized as follows. In Sec.~\ref{sec2}, we briefly summarize the Higgs sector of the $\mu\nu$SSM, including the Higgs boson mass corrections. We present the diagonalization of the neutral scalar mass matrix analytically in Sec.~\ref{sec3}. The numerical analyses are given in Sec.~\ref{sec-num}, and Sec.~\ref{sec-sum} provides a summary.  The tedious formulas are collected in the Appendixes.

\section{The Higgs sector\label{sec2}}
\indent
The Higgs sector of the $\mu\nu$SSM contains the usual two Higgs doublets with the left- and right-handed sneutrinos: $\hat H_d^T = \Big( {\hat H_d^0,\hat H_d^ - } \Big)$, $\hat H_u^T = \Big( {\hat H_u^ + ,\hat H_u^0} \Big)$,  $\hat{\nu}_i$ and  $\hat{\nu}_i^c$. Once EWSB, the neutral scalars have the VEVs:
\begin{eqnarray}
\langle H_d^0 \rangle = \upsilon_d , \qquad \langle H_u^0 \rangle = \upsilon_u , \qquad
\langle \tilde \nu_i \rangle = \upsilon_{\nu_i} , \qquad \langle \tilde \nu_i^c \rangle = \upsilon_{\nu_i^c} .
\end{eqnarray}
One can define the neutral scalars as
\begin{eqnarray}
&&H_d^0=\frac{1}{\sqrt{2}} \Big(h_d + i P_d \Big) + \upsilon_d, \qquad\; \tilde \nu_i = \frac{1}{\sqrt{2}} \Big((\tilde \nu_i)^\Re + i (\tilde \nu_i)^\Im \Big) + \upsilon_{\nu_i},  \nonumber\\
&&H_u^0=\frac{1}{\sqrt{2}} \Big(h_u + i P_u \Big) + \upsilon_u, \qquad \tilde \nu_i^c = \frac{1}{\sqrt{2}} \Big((\tilde \nu_i^c)^\Re + i (\tilde \nu_i^c)^\Im \Big) + \upsilon_{\nu_i^c},
\end{eqnarray}

Considering that the neutrino oscillation data constrain neutrino Yukawa couplings $Y_{\nu_i} \sim \mathcal{O}(10^{-7})$ and left-handed sneutrino VEVs $\upsilon_{\nu_i} \sim \mathcal{O}(10^{-4}{\rm{GeV}})$~\cite{mnSSM,mnSSM1,mnSSM2,ref-zhang1,meu-m,meu-m1,meu-m2,meu-m3,
neu-zhang1,neu-zhang2}, in the following we could reasonably neglect the small terms including $Y_{\nu}$ or $\upsilon_{\nu_i}$ in the Higgs sector. Then, the superpotential in Eq.~(\ref{eq-W}) approximately leads to the tree-level neutral scalar (Higgs) potential:
\begin{eqnarray}
V^0=V_F+V_D+V_{soft},
\end{eqnarray}
with
\begin{eqnarray}
&&V_F=\lambda_i \lambda_i^* H_d^0 H_d^{0*} H_u^0 H_u^{0*}+\lambda_i \lambda_j^* \tilde\nu_i^c \tilde\nu_j^{c*} (H_d^0 H_d^{0*}+ H_u^0 H_u^{0*})\nonumber\\
&&\hspace{1.0cm} + \, \kappa_{ijk}\kappa_{ljm}^* \tilde\nu_i^c \tilde\nu_k^{c} \tilde\nu_l^{c*} \tilde\nu_m^{c*} - ( \kappa_{ijk}\lambda_j^* \tilde\nu_i^c \tilde\nu_k^{c} H_d^{0*} H_u^{0*}+ {\rm{H.c.}}),\\
&&V_D=\frac{G^2}{8}(\tilde\nu_i \tilde\nu_i^{*} +H_d^0 H_d^{0*}- H_u^0 H_u^{0*})^2,\\
&&V_{soft}=m_{{H_d}}^{2} H_d^0 H_d^{0*} + m_{{H_u}}^2 H_u^0 H{_u^{0*}}
+ m_{{{\tilde L}_{ij}}}^2\tilde \nu_i \tilde\nu_j^{*} + m_{\tilde \nu_{ij}^c}^2\tilde \nu{_i^{c}}\tilde \nu_j^{c*}\nonumber\\
&&\hspace{1.3cm} - \,\Big((A_{\lambda}\lambda)_i \nu{_i^{c}} H_d^0 H_u^{0} - \frac{1}{3} (A_{\kappa}\kappa)_{ijk} \tilde\nu{_i^{c}}\tilde \nu{_j^{c}}\tilde \nu{_k^{c}} + {\rm{H.c.}}\Big),
\end{eqnarray}
where $G^2=g_1^2+g_2^2$ and $g_1 c_{_W} =g_2 s_{_W}=e$, $V_F$ and $V_D$ are the usual $F$ and $D$ terms derived from the superpotential, and $V_{soft}$ denotes the soft supersymmetry breaking terms. For simplicity, we will assume that all parameters in the potential are real in the following.

With effective potential methods~\cite{Hi-1,Hi-2,Hi-3,Hi-4,Hi-5,Hi-6,Hi-7,Hi-8,Hi-alpha,Hi-9,Hi-10,Hi-11,Hi-12,Hi-13,Hi-14,Hi-15}, the one-loop effective potential can be given by
\begin{eqnarray}
&&V^1=\frac{1}{32\pi^2}\Big\{ \sum\limits_{\tilde f} N_f m_{\tilde f}^4 \Big( \log \frac{m_{\tilde f}^2}{Q^2} -\frac{3}{2}\Big) -2 \sum\limits_{f= t, b,\tau} N_f  m_{f}^4 \Big( \log \frac{m_{f}^2}{Q^2} -\frac{3}{2}\Big) \Big\},
\end{eqnarray}
where, $Q$ denotes the renormalization scale, $N_t=N_b=3$ and $N_\tau=1$, $\tilde f=\tilde t_{1,2},\tilde b_{1,2},\tilde \tau_{1,2}$. The masses of the third fermions $f= t, b,\tau$ and corresponding supersymmetric partners $\tilde f=\tilde t_{1,2},\tilde b_{1,2},\tilde \tau_{1,2}$ in the $\mu\nu$SSM are collected in Appendix~\ref{app1}. Including the one-loop effective potential, the Higgs potential is written as
\begin{eqnarray}
V=V^0+V^1.
\end{eqnarray}
Through the Higgs potential, we will calculate the minimization conditions of the potential and the Higgs masses in the following.

Minimizing the Higgs potential, we can obtain the minimization conditions of the potential, linking the soft mass parameters to the VEVs of the neutral scalar fields:
\begin{eqnarray}
&&m_{{H_d}}^2= -\Delta T_{H_d} + ((A_\lambda \lambda)_i \upsilon_{\nu_i^c} + {\lambda _j}{\kappa _{ijk}}\upsilon_{\nu_i^c} \upsilon_{\nu_k^c} )\tan\beta \nonumber\\
&&\hspace{1.4cm} - \, ({\lambda _i}{\lambda _j}\upsilon_{\nu_i^c}\upsilon_{\nu_j^c}  + {\lambda _i}{\lambda _i}\upsilon_u^2)  + \frac{G^2}{4}( \upsilon_u^2 - \upsilon_d^2),\\
&&m_{{H_u}}^2= -\Delta T_{H_u} + ((A_\lambda \lambda)_i \upsilon_{\nu_i^c} +  {\lambda _j}{\kappa _{ijk}}\upsilon_{\nu_i^c} \upsilon_{\nu_k^c})\cot\beta\nonumber\\
&&\hspace{1.4cm} - \, ({\lambda _i}{\lambda _j}\upsilon_{\nu_i^c}\upsilon_{\nu_j^c} + {\lambda _i}{\lambda _i}\upsilon_d^2)  + \frac{{G^2}}{4}(\upsilon_d^2 - \upsilon_u^2) , \\
&&m_{\tilde \nu_{ij}^c}^2 \upsilon_{\nu_j^c}= -\Delta T_{\tilde \nu_{ij}^c} \upsilon_{\nu_j^c} +  (A_\lambda \lambda)_i{\upsilon_d}{\upsilon_u} - {( A_\kappa \kappa)}_{ijk} \upsilon_{\nu_j^c} \upsilon_{\nu_k^c}  +  2{\lambda _j}{\kappa _{ijk}}\upsilon_{\nu_k^c}{\upsilon_d}{\upsilon_u} \nonumber\\
&&\hspace{1.9cm} - \, 2{\kappa _{lim}}{\kappa _{ljk}} \upsilon_{\nu_m^c} \upsilon_{\nu_j^c} \upsilon_{\nu_k^c}  -  {\lambda _i}{\lambda _j}\upsilon_{\nu_j^c}(\upsilon_d^2  + \upsilon_u^2) , \quad(i=1,2,3)
\end{eqnarray}
where, as usual, $\tan\beta ={\upsilon_u}/{\upsilon_d}$. $\Delta T_{H_d}$, $\Delta T_{H_u}$, and $\Delta T_{\tilde \nu_{ij}^c} \upsilon_{\nu_j^c}$ come from one-loop corrections to the minimization conditions, which are taken in Appendix~\ref{app2}. Here, neglecting the small terms including $Y_{\nu}$ or $\upsilon_{\nu_i}$ in the Higgs sector, we do not give the minimization conditions of the potential about the left-handed sneutrino VEVs, which can be used to constrain $\upsilon_{\nu_i}$~\cite{meu-m,neu-zhang2}.

From the Higgs potential, one can derive the $8\times8$ mass matrices for the CP-even neutral scalars ${S'^T} = ({h_d},{h_u},{(\tilde \nu_i^c)^\Re},{({\tilde \nu_i})^\Re})$ and the CP-odd neutral scalars ${P'^T} = ({P_d},{P_u},{(\tilde \nu_i^c)^\Im},{({\tilde \nu_i})^\Im})$ in the unrotated basis. Ignoring the small terms including $Y_{\nu}$ or $\upsilon_{\nu_i}$, the $5\times5$ mass submatrix for Higgs
doublets and right-handed sneutrinos is basically decoupled from the $3\times3$ left-handed sneutrinos mass submatrix. The $3\times3$ left-handed sneutrino mass submatrix is $\Big(m_{\tilde L_{ij}}^2 + \frac{G^2}{4}(\upsilon_d^2 - \upsilon_u^2 )\delta_{ij}\Big)_{3\times3}$, which is dominated by the soft mass $m_{\tilde L_{ij}}^2$. Through the Higgs potential, the $5\times5$ mass submatrix for Higgs doublets and right-handed sneutrinos in the CP-even sector can be derived as
\begin{eqnarray}
M_S^2 = \left( {\begin{array}{*{20}{c}}
   M_{H}^2 & M_{X}^2   \\
   \Big(M_{X}^{2}\Big)^T & M_{R}^2   \\
\end{array}} \right),
\end{eqnarray}
where $M_{H}^2$ denotes the $2\times2$ mass submatrix for Higgs doublets, $M_{R}^2$ is the $3\times3$ mass submatrix for right-handed sneutrinos and  $M_{X}^2$ represents the $2\times3$ mass submatrix for the mixing of Higgs doublets and right-handed sneutrinos.

In detail, the $2\times2$ mass submatrix $M_{H}^2$ can be written by
\begin{eqnarray}
M_H^2 = \left( {\begin{array}{*{20}{c}}
   M_{h_d h_d}^2+\Delta_{11} & M_{h_d h_u}^2+\Delta_{12}   \\
   M_{h_d h_u}^2+\Delta_{12} & M_{h_u h_u}^2+\Delta_{22}   \\
\end{array}} \right),
\end{eqnarray}
with the tree-level contributions as
\begin{eqnarray}
&&M_{h_d h_u}^2 =  -\Big[m_A^2+\Big(1-4\lambda_i \lambda_i s_{_W}^2 c_{_W}^2/e^2 \Big)m_Z^2\Big] \sin \beta \cos \beta, \\
&&M_{h_d h_d}^2 =  m_A^2 \sin^2 \beta+m_Z^2 \cos^2 \beta, \\
&&M_{h_u h_u}^2 =  m_A^2 \cos^2 \beta+m_Z^2 \sin^2 \beta,
\end{eqnarray}
and the neutral pseudoscalar mass squared as
\begin{eqnarray}
m_A^2\simeq \frac{2}{{\sin2\beta}}\Big[(A_\lambda \lambda)_i \upsilon_{\nu_i^c}+ {\lambda _k}{\kappa _{ijk}}\upsilon_{\nu_i^c} \upsilon_{\nu_j^c}\Big] .
\end{eqnarray}
Comparing with the MSSM, $M_{h_d h_u}^2$ has an additional term $(4\lambda_i \lambda_i s_{W}^2 c_{W}^2/e^2 )m_Z^2 \sin \beta \cos \beta$, which can give a new contribution to the lightest Higgs boson mass. The radiative corrections $\Delta_{11}$, $\Delta_{12}$, and $\Delta_{22}$ from the third fermions ${f}={t},{b},{\tau}$ and their superpartners can be found in Ref.~\cite{ref-zhang2}, which agree with the results of the MSSM~\cite{Hi-1,Hi-2,Hi-3,Hi-4,Hi-5,Hi-6,Hi-7,Hi-8,Hi-alpha,Hi-9,Hi-10,Hi-11,Hi-12,Hi-13}. Here, the radiative corrections from the top quark and its superpartners include the two-loop leading-log effects, which can obviously affect the mass of the lightest Higgs boson.

Furthermore, the $2\times3$ mixing mass submatrix $M_{X}^2$ is
\begin{eqnarray}
M_X^2 = \left( {\begin{array}{*{20}{c}}
   \Big(M_{h_d (\tilde \nu_i^c)^\Re}^2+\Delta_{1(2+i)}\Big)_{1\times3}    \\
   \Big(M_{h_u (\tilde \nu_i^c)^\Re}^2+\Delta_{2(2+i)}\Big)_{1\times3}    \\
\end{array}} \right),
\end{eqnarray}
where
\begin{eqnarray}
&&M_{h_d (\tilde \nu_i^c)^\Re}^2 = \Big[ 2\lambda_i \lambda_j \upsilon_{\nu_j^c} \cot\beta  - \Big( (A_\lambda \lambda)_i + 2 \lambda_k \kappa_{ijk} \upsilon_{\nu_j^c} \Big) \Big] \upsilon_u \,, \\
&&M_{h_u (\tilde \nu_i^c)^\Re}^2 = \Big[ 2\lambda_i \lambda_j \upsilon_{\nu_j^c} \tan\beta  - \Big((A_\lambda \lambda)_i + 2 \lambda_k \kappa_{ijk}\upsilon_{\nu_j^c} \Big)  \Big]  \upsilon_d \,,
\end{eqnarray}
and the radiative corrections from the third fermions ${f}={t},{b},{\tau}$ and their superpartners are
\begin{eqnarray}
&&\Delta_{1(2+i)} =  \lambda_i \upsilon_u \Delta_{1R} \,, \qquad \Delta_{2(2+i)} = \lambda_i \upsilon_d \Delta_{2R} \, ,\\
&&\Delta_{1R} = \frac{G_F}{2\sqrt{2}\pi^2}  \Big\{ \frac{3{m_{t}^4}}{\sin^2\beta} {\mu (A_{t}-\mu\cot\beta)^2\over {\tan\beta (m_{\tilde{t}_1}^2-m_{\tilde{t}_2}^2)}^2}
g(m_{\tilde{t}_1}^2,m_{\tilde{t}_2}^2)
\nonumber\\
&&\hspace{1.3cm}
+ \frac{3{m_{b}^4}}{\cos^2\beta} {(-A_{b}+\mu\tan\beta)\over (m_{\tilde{b}_1}^2-m_{\tilde{b}_2}^2)} \Big[\log{m_{\tilde{b}_1}^2\over m_{\tilde{b}_2}^2} +{A_{b}(A_{b}-\mu\tan\beta)\over {(m_{\tilde{b}_1}^2-m_{\tilde{b}_2}^2)}}
g(m_{\tilde{b}_1}^2,m_{\tilde{b}_2}^2)\Big] \nonumber\\
&&\hspace{1.3cm}
+ \frac{{m_{\tau}^4}}{\cos^2\beta}{(-A_{\tau}+\mu\tan\beta)\over (m_{\tilde{\tau}_1}^2-m_{\tilde{\tau}_2}^2)} \Big[ \log{m_{\tilde{\tau}_1}^2\over m_{\tilde{\tau}_2}^2} +{A_{\tau}(A_{\tau}-\mu\tan\beta)\over {(m_{\tilde{\tau}_1}^2-m_{\tilde{\tau}_2}^2)}}
g(m_{\tilde{\tau}_1}^2,m_{\tilde{\tau}_2}^2)\Big] \Big\} \, ,\\
&&\Delta_{2R} = \frac{G_F}{2\sqrt{2}\pi^2} \Big\{ \frac{3{m_{t}^4}}{\sin^2\beta} {(-A_{t}+\mu\cot\beta)\over (m_{\tilde{t}_1}^2-m_{\tilde{t}_2}^2)}\Big[ \log{m_{\tilde{t}_1}^2\over m_{\tilde{t}_2}^2}
+{A_{t}(A_{t}-\mu\cot\beta)\over {(m_{\tilde{t}_1}^2-m_{\tilde{t}_2}^2)}}
g(m_{\tilde{t}_1}^2,m_{\tilde{t}_2}^2)\Big]
\nonumber\\
&&\hspace{1.3cm}+\frac{3{m_{b}^4}}{\cos^2\beta}{\mu (A_{b}-\mu\tan\beta)^2\over {\cot\beta(m_{\tilde{b}_1}^2-m_{\tilde{b}_2}^2)}^2}g(m_{\tilde{b}_1}^2,m_{\tilde{b}_2}^2) \nonumber\\
&&\hspace{1.3cm}
+ \frac{{m_{\tau}^4}}{\cos^2\beta}{\mu(A_{\tau}-\mu\tan\beta)^2\over {\cot\beta(m_{\tilde{\tau}_1}^2-m_{\tilde{\tau}_2}^2)}^2}
g(m_{\tilde{\tau}_1}^2,m_{\tilde{\tau}_2}^2) \Big\}\, ,
\end{eqnarray}
with $\mu=\lambda_i \upsilon_{\nu_i^c}$, $g(m_1^2,m_2^2)=2-{m_1^2+m_2^2\over m_1^2-m_2^2}\log{m_1^2\over m_2^2}$. Here, we can know that the radiative corrections to the mixing are proportional to the parameters $\lambda_i$.

Similarly, one can derive the $3\times3$ mass submatrix for the right-handed sneutrinos:
\begin{eqnarray}
M_R^2 = \left( {\begin{array}{*{20}{c}}
   M_{(\tilde \nu_i^c)^\Re (\tilde \nu_j^c)^\Re }^2+\Delta_{(2+i)(2+j)} \\
\end{array}} \right)_{3\times3},
\end{eqnarray}
with
\begin{eqnarray}
&&M_{(\tilde \nu_i^c)^\Re (\tilde \nu_j^c)^\Re }^2 =  m_{\tilde \nu_{ij}^c}^2 + 2 {(A_\kappa \kappa)}_{ijk} \upsilon_{\nu_k^c} - 2\lambda_k \kappa_{ijk} \upsilon_d \upsilon_u + \lambda_i \lambda_j ( \upsilon_d^2 + \upsilon_u^2) \nonumber\\
&&\hspace{2.5cm} +\:(2\kappa_{ijk}\kappa_{lmk}+4\kappa_{ilk}\kappa_{jmk}) \upsilon_{\nu_l^c}\upsilon_{\nu_m^c} \,,
\end{eqnarray}
and the corrections from the third fermions and their superpartners are
\begin{eqnarray}
&&\Delta_{(2+i)(2+j)} =  \lambda_i \lambda_j \Delta_{RR} \,, \\
&&\Delta_{RR} = \frac{G_F}{2\sqrt{2}\pi^2}  \Big\{ \frac{3{m_{t}^4}}{\sin^2\beta} {\upsilon_d^2 (A_{t}-\mu\cot\beta)^2\over {(m_{\tilde{t}_1}^2-m_{\tilde{t}_2}^2)}^2}
g(m_{\tilde{t}_1}^2,m_{\tilde{t}_2}^2)
\nonumber\\
&&\hspace{1.45cm}
+\frac{3{m_{b}^4}}{\cos^2\beta}{\upsilon_u^2 (A_{b}-\mu\tan\beta)^2\over {(m_{\tilde{b}_1}^2-m_{\tilde{b}_2}^2)}^2}g(m_{\tilde{b}_1}^2,m_{\tilde{b}_2}^2) \nonumber\\
&&\hspace{1.45cm}
+ \frac{{m_{\tau}^4}}{\cos^2\beta}{\upsilon_u^2 (A_{\tau}-\mu\tan\beta)^2\over {(m_{\tilde{\tau}_1}^2-m_{\tilde{\tau}_2}^2)}^2}
g(m_{\tilde{\tau}_1}^2,m_{\tilde{\tau}_2}^2) \Big\}\, .
\end{eqnarray}
Here, the radiative corrections to the mass submatrix for right-handed sneutrinos are proportional to  $\lambda_i\lambda_j$.

\section{Diagonalization of the mass matrix\label{sec3}}
\indent
The mass squared matrix $M_{H}^2$ which contains the radiative corrections can be diagonalized as
\begin{eqnarray}
U_H^T M_{H}^2 U_H = {\rm{diag}} \Big(m_{H_1}^2,m_{H_2}^2\Big),
\end{eqnarray}
by the $2\times2$ unitary matrix $U_H$,
\begin{eqnarray}
U_H=
\left(\begin{array}{*{20}{c}}
-\sin \alpha & \cos \alpha\\
\cos \alpha & \sin \alpha
\end{array}\right).
\end{eqnarray}
Here, the neutral doubletlike Higgs mass squared eigenvalues $m_{{H_{1,2}}}^2$ can be derived,
\begin{eqnarray}
m_{{H_{1,2}}}^2={1\over 2}\Big[{\rm{Tr}}M_{H}^2 \mp\sqrt{({{\rm{Tr}}M_{H}^2})^2-4{\rm{Det}}M_{H}^2}\Big],
\end{eqnarray}
where ${\rm{Tr}}M_{H}^2={M_{H}^2}_{11}+{M_{H}^2}_{22}$, ${\rm{Det}}{M_{H}^2} = {M_{H}^2}_{11}{M_{H}^2}_{22}-({M_{H}^2}_{12})^2$.
The mixing angle $\alpha$ can be determined by~\cite{Hi-alpha}
\begin{eqnarray}
&&\sin 2\alpha =\frac{2{M_{H}^2}_{12}}{\sqrt{({\rm{Tr}}{M_{H}^2})^2-4{\rm{Det}}{M_{H}^2}}},\nonumber\\
&&\cos 2\alpha =\frac{{M_{H}^2}_{11}-{M_{H}^2}_{22}}{\sqrt{({\rm{Tr}}{M_{H}^2})^2-4{\rm{Det}}{M_{H}^2}}},
\end{eqnarray}
which reduce to $-\sin 2\beta$ and $-\cos 2\beta$, respectively, in the large $m_A$ limit. The convention is that $\pi/4\leq\beta<\pi/2$ for $\tan \beta\geq1$, while $-\pi/2<\alpha<0$. In the large $m_A$ limit, $\alpha=-\pi/2+\beta$.

In the large $m_A$ limit, the light neutral doubletlike Higgs mass is approximately given as
\begin{eqnarray}
m_{H_{1}}^2 \simeq m_Z^2 \cos^2 2\beta + \frac{2 \lambda_i \lambda_i s_{_W}^2 c_{_W}^2}{ e^2} m_Z^2 \sin^2 2\beta+\bigtriangleup m_{H_{1}}^2.
\label{mH1-app}
\end{eqnarray}
Comparing with the MSSM, the $\mu\nu{\rm SSM}$ gets an additional term $\frac{2 \lambda_i \lambda_i s_{_W}^2 c_{_W}^2}{ e^2} m_Z^2 \sin^2 2\beta$~\cite{mnSSM1}.
Here, the radiative corrections $\bigtriangleup m_{H_{1}}^2$ can be computed more precisely by some public tools, for example, FeynHiggs~\cite{FeynHiggs-1,FeynHiggs-2,FeynHiggs-3,FeynHiggs-4,FeynHiggs-5,FeynHiggs-6,FeynHiggs-7,FeynHiggs-8}, SOFTSUSY~\cite{SOFTSUSY-1,SOFTSUSY-2,SOFTSUSY-3}, SPheno~\cite{SPheno-1,SPheno-2}, and so on. In the following numerical section, we will use the FeynHiggs-2.13.0 to calculate the radiative corrections for the Higgs boson mass about the MSSM part.

To further deal with the mass submatrix $M_{R}^2$ and $M_{X}^2$, in the following we choose the usual minimal scenario for the parameter space:
\begin{eqnarray}
&&\lambda _i = \lambda , \quad
({A_\lambda }\lambda )_i = {A_\lambda }\lambda, \quad
\upsilon_{\nu_i^c}=\upsilon_{\nu^c},
\nonumber\\
&&{\kappa _{ijk}} = \kappa {\delta _{ij}}{\delta _{jk}}, \quad
{({A_\kappa }\kappa )_{ijk}} = {A_\kappa }\kappa {\delta _{ij}}{\delta _{jk}}, \quad
m_{\tilde \nu_{ij}^c}^2 = m_{{{\tilde \nu_i}^c}}^2{\delta _{ij}},
\label{MSPS}
\end{eqnarray}
Then, the $3\times3$ mass submatrix for CP-even right-handed sneutrinos can be simplified as
\begin{eqnarray}
M_R^2 = \left( {\begin{array}{*{20}{c}}
   X_{R} & y_{_R} & y_{_R}   \\
   y_{_R} & X_{R} & y_{_R}   \\
   y_{_R} & y_{_R} & X_{R}   \\
\end{array}} \right),
\end{eqnarray}
with
\begin{eqnarray}
&&X_{R} = (A_\kappa+4\kappa\upsilon_{\nu^c})\kappa\upsilon_{\nu^c} +A_\lambda \lambda \upsilon_d \upsilon_u/\upsilon_{\nu^c} + \lambda^2 \Delta_{RR}\,,\\
&&y_{_R} = \lambda^2 ( \upsilon^2+\Delta_{RR})\,,
\end{eqnarray}
where $\upsilon^2=\upsilon_d^2+\upsilon_u^2$.
Here the radiative corrections keep the dominating contributions which are proportional to $m_f^4$ ($f=t,b,\tau$).
Through the $3\times3$ unitary matrix $U_R$,
\begin{eqnarray}
U_R=
\left(\begin{array}{*{20}{c}}
   \frac{1}{\sqrt{3}} & 0 & -\frac{2}{\sqrt{6}}   \\
   \frac{1}{\sqrt{3}} & -\frac{1}{\sqrt{2}} & \frac{1}{\sqrt{6}}   \\
   \frac{1}{\sqrt{3}} & \frac{1}{\sqrt{2}} & \frac{1}{\sqrt{6}}
\end{array}\right),
\end{eqnarray}
the mass squared matrix $M_{R}^2$ can be diagonalized as
\begin{eqnarray}
U_R^T M_{R}^2 U_R = {\rm{diag}} \Big(m_{R_1}^2,m_{R_2}^2,m_{R_3}^2\Big),
\end{eqnarray}
with
\begin{eqnarray}
\label{mR1}
&&m_{R_1}^2=X_{R}+2y_{_R}= (A_\kappa+4\kappa\upsilon_{\nu^c})\kappa\upsilon_{\nu^c} +A_\lambda \lambda \upsilon_d \upsilon_u/\upsilon_{\nu^c} + \lambda^2 (2 \upsilon^2+3\Delta_{RR}),\\
&&m_{R_2}^2=m_{R_3}^2=X_{R}-y_{_R}= (A_\kappa+4\kappa\upsilon_{\nu^c})\kappa\upsilon_{\nu^c} +A_\lambda \lambda \upsilon_d \upsilon_u/\upsilon_{\nu^c} - \lambda^2 \upsilon^2.
\end{eqnarray}
The radiative corrections are proportional to $\lambda^2$, which will be tamped down as $\lambda \sim \mathcal{O}(0.1)$. Then the masses squared of the CP-even right-handed sneutrinos can be approximated by
\begin{eqnarray}
m_{S_R}^2 \approx m_{R_1}^2 \approx  m_{R_2}^2 =  m_{R_3}^2 \approx (A_\kappa+4\kappa\upsilon_{\nu^c})\kappa\upsilon_{\nu^c} +A_\lambda \lambda \upsilon_d \upsilon_u/\upsilon_{\nu^c}\, .
\label{mSR}
\end{eqnarray}
Due to $\upsilon_{\nu^c} \gg \upsilon_{u,d}$, the main contribution to the mass squared is the first term as $\kappa$ is large. Additionally, the masses squared of the CP-odd right-handed sneutrinos $m_{P_R}^2$ can be approximated as
\begin{eqnarray}
m_{P_R}^2 \approx -3A_\kappa\kappa\upsilon_{\nu^c}+(4\kappa+A_\lambda /\upsilon_{\nu^c})\lambda \upsilon_d \upsilon_u\, ,
\label{mPR}
\end{eqnarray}
where the first term is the leading contribution.  Therefore, one can use the approximate relation,
\begin{eqnarray}
-4\kappa\upsilon_{\nu^c}\lesssim A_\kappa \lesssim 0\, ,
\label{tachyon}
\end{eqnarray}
to avoid the tachyons.

In the minimal scenario for the parameter space presented in Eq.~(\ref{MSPS}), the $2\times3$ mixing mass submatrix $M_{X}^2$ is simplified as
\begin{eqnarray}
M_X^2 = \left( {\begin{array}{*{20}{c}}
   M_{X_1}^2 &  M_{X_1}^2 & M_{X_1}^2   \\
   M_{X_2}^2 &  M_{X_2}^2 & M_{X_2}^2   \\
\end{array}} \right),
\end{eqnarray}
where
\begin{eqnarray}
&&M_{X_1}^2 = \lambda \upsilon \sin\beta \Big[ 2\upsilon_{\nu^c} ( 3\lambda \cot\beta - \kappa )- A_\lambda +  \Delta_{1R} \Big] \,, \\
&&M_{X_2}^2 = \lambda \upsilon \cos\beta \Big[ 2\upsilon_{\nu^c} ( 3\lambda \tan\beta - \kappa )- A_\lambda +   \Delta_{2R}  \Big] \,.
\end{eqnarray}
Then, we do the calculation:
\begin{eqnarray}
\left( {\begin{array}{*{20}{c}}
   U_{H}^T & 0   \\
   0 &  U_{R}^T   \\
\end{array}} \right)
\left( {\begin{array}{*{20}{c}}
   M_{H}^2 & M_{X}^2   \\
   \Big(M_{X}^{2}\Big)^T & M_{R}^2   \\
\end{array}} \right)
\left( {\begin{array}{*{20}{c}}
   U_{H} & 0   \\
   0 &  U_{R}   \\
\end{array}} \right)
= {\cal H}
\oplus \left( {\begin{array}{*{20}{c}}
   m_{R_2}^2 & 0   \\
   0 &  m_{R_3}^2   \\
\end{array}} \right),
\end{eqnarray}
with
\begin{eqnarray}
{\cal H}= \left( {\begin{array}{*{20}{c}}
   m_{H_1}^2 & 0 & A_{X_1}^2   \\
   0 & m_{H_2}^2 & A_{X_2}^2   \\
   A_{X_1}^2 & A_{X_2}^2 & m_{R_1}^2   \\
\end{array}} \right),
\end{eqnarray}
where
\begin{eqnarray}
&&A_{X_1}^2 = \sqrt{3} (-M_{X_1}^2 \sin\alpha + M_{X_2}^2 \cos\alpha) \,, \\
&&A_{X_2}^2 = \sqrt{3} (M_{X_1}^2 \cos\alpha + M_{X_2}^2 \sin\alpha) \,.
\end{eqnarray}
In the large $m_A$ limit, $\alpha=-\pi/2+\beta$. Then, one can have the following approximate expressions:
\begin{eqnarray}
\label{AX1}
&&A_{X_1}^2 \simeq \sqrt{3} \lambda \upsilon \sin2\beta \Big[ 2\upsilon_{\nu^c} \Big( \frac{3\lambda}{\sin2\beta}  - \kappa \Big)- A_\lambda + \frac{1}{2} (\Delta_{1R} + \Delta_{2R}) \Big] \,, \\
\label{AX2}
&&A_{X_2}^2 \simeq \sqrt{3} \lambda \upsilon \Big[ ( 2\kappa \upsilon_{\nu^c}+ A_\lambda ) \cos2\beta + \Delta_{1R}\sin^2\beta  - \Delta_{2R}\cos^2\beta \Big] \,.
\end{eqnarray}

If $A_{X_1}^2=0$, the mixing of Higgs doublets and right-handed sneutrinos will not affect the lightest Higgs boson mass~\cite{mnSSM1}; namely, one can adopt the relation
\begin{eqnarray}
A_\lambda = 2\upsilon_{\nu^c} \Big( \frac{3\lambda}{\sin2\beta}  - \kappa \Big) + \frac{1}{2} (\Delta_{1R} + \Delta_{2R})  \,,
\label{Alambda}
\end{eqnarray}
which is analogous to the NMSSM~\cite{ref-NMSSM1,ref-NMSSM2}. To relax the conditions, if $A_\lambda$ is around the value in Eq.~(\ref{Alambda}), the contribution to the lightest Higgs boson mass from the mixing could also be neglected approximately. In the case $A_{X_1}^2\approx0$, the mass of the lightest Higgs boson is just $m_{H_1}$, which shows, approximately, in Eq.~(\ref{mH1-app}).

If $A_{X_1}^2\neq0$, we need to diagonalize the $3\times3$ mass matrix ${\cal H}$ further:
\begin{eqnarray}
U_X^T {\cal H} U_X = {\rm{diag}} \Big(m_h^2,m_H^2,m_{S_3}^2\Big),
\end{eqnarray}
where the eigenvalues $m_h^2,m_H^2,m_{S_3}^2$ and the unitary matrix $U_X$ can be concretely seen in Appendix~\ref{app3}. Then, the lightest Higgs boson mass is exactly $m_h^2$. In the large $m_A$ limit, $m_{H_2}\simeq m_A$, one can have the lightest Higgs boson mass squared approximately as
\begin{eqnarray}
m_h^2 \simeq \frac{1}{2} \Big\{m_{H_1}^2+m_{R_1}^2-\frac{ (A_{X_2}^2)^2}{m_{H_2}^2} -\sqrt{\Big[m_{R_1}^2-m_{H_1}^2-\frac{(A_{X_2}^2)^2}{m_{H_2}^2}\Big]^2 +4(A_{X_1}^2)^2}\Big\}.
\label{mh-app}
\end{eqnarray}
The approximate expression works well, which can be easily checked in the numerical calculation. When $m_{H_2}$ and $m_{R_1}$ are all large, Eq.~(\ref{mh-app}) could be approximated by
\begin{eqnarray}
m_h^2 \approx  m_{H_1}^2-\frac{(A_{X_1}^2)^2}{m_{R_1}^2} = m_{H_1}^2 \Big[1-\frac{(A_{X_1}^2)^2}{m_{R_1}^2 m_{H_1}^2}\Big].
\label{mh-app1}
\end{eqnarray}
In the numerical analysis, we can define the quantity
\begin{eqnarray}
\xi_h = \frac{(A_{X_1}^2)^2}{m_{R_1}^2 m_{H_1}^2}\,
\label{xih}
\end{eqnarray}
to analyze how the mixing affects the mass of the lightest Higgs boson.

One can diagonalize the $5\times5$ mass submatrix for Higgs doublets and right-handed sneutrinos in the CP-even sector:
\begin{eqnarray}
R_S^T M_{S}^2 R_S = {\rm{diag}} \Big(m_{S_1}^2,m_{S_2}^2,m_{S_3}^2,m_{S_4}^2,m_{S_5}^2\Big),
\end{eqnarray}
with $m_{S_1}=m_h,\:m_{S_2}=m_H,\:m_{S_4}=m_{R_2}=m_{S_5}=m_{R_3}$, and the $5\times5$ unitary matrix $R_S$
\begin{eqnarray}
R_S=
\left( {\begin{array}{*{20}{c}}
   U_{H} & 0   \\
   0 &  U_{R}   \\
\end{array}} \right)
\left( {\begin{array}{*{20}{c}}
   U_X & 0   \\
   0 &  I_{2\times2}   \\
\end{array}} \right),
\end{eqnarray}
where $I_{2\times2}$ denotes the ${2\times2}$ unit matrix.

\section{Numerical analysis\label{sec-num}}
\indent
In this section, we will do the numerical analysis for the masses of the Higgs bosons. First, we choose the values of the parameter space.
For the relevant parameters in the SM, we choose
\begin{eqnarray}
&&\alpha_s(m_{Z})=0.118,\qquad   m_{Z}=91.188\;{\rm GeV}, \qquad m_{W}=80.385\;{\rm GeV},
\nonumber\\
&&m_t=173.2\;{\rm GeV},\qquad   m_b=4.66\;{\rm GeV}, \qquad\: m_{\tau}=1.777\;{\rm GeV}.
\end{eqnarray}
The other SM parameters can be seen in Ref.~\cite{PDG} from the Particle Data Group.
Here, we choose a suitable ${A_{\kappa}}=-500\;{\rm GeV}$ to avoid the tachyons easily, through Eq.~(\ref{tachyon}). Considering the direct search for supersymmetric particles~\cite{PDG}, we could reasonably choose $M_2=2M_1=800\;{\rm GeV}$, $M_3=2\;{\rm TeV}$, $m_{{\tilde Q}_3}=m_{{\tilde U}_3}=m_{{\tilde D}_3}=2\;{\rm TeV}$, $m_{{\tilde L}_{3}}=m_{{\tilde E}_{3}}=1\;{\rm TeV}$, $A_{b}=A_{\tau}=1\;{\rm TeV}$, and $A_{t}=2.5\;{\rm TeV}$ for simplicity. As key parameters, $m_{{\tilde Q}_3}$, $m_{{\tilde U}_3}$, $A_t$ and the gaugino mass parameters affect the radiative corrections to the lightest Higgs mass. Therefore, one can take the proper values for $m_{{\tilde Q}_3}$, $m_{{\tilde U}_3}$, $A_t$ and the gaugino mass parameters to keep the lightest Higgs mass around $125$ GeV.

In the following, we will analyze how the mixing of Higgs doublets and right-handed sneutrinos affects the lightest Higgs boson mass. Through $A_{X_1}^2$ in Eq.~(\ref{AX1}), one knows that the parameters which affect the lightest Higgs boson mass from the mixing will be $\lambda,\,\tan\beta,\,\kappa,\,A_\lambda$, and $\upsilon_{\nu^c}$. Here, we specify that the parameter $\mu=3\lambda\upsilon_{\nu^c}$, which is dominated by the parameters $\lambda$ and $\upsilon_{\nu^c}$.

\begin{figure}
\setlength{\unitlength}{1mm}
\centering
\begin{minipage}[c]{0.45\textwidth}
\includegraphics[width=2.9in]{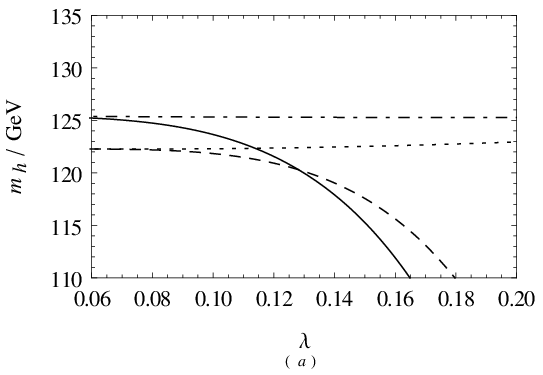}
\end{minipage}%
\begin{minipage}[c]{0.45\textwidth}
\includegraphics[width=2.9in]{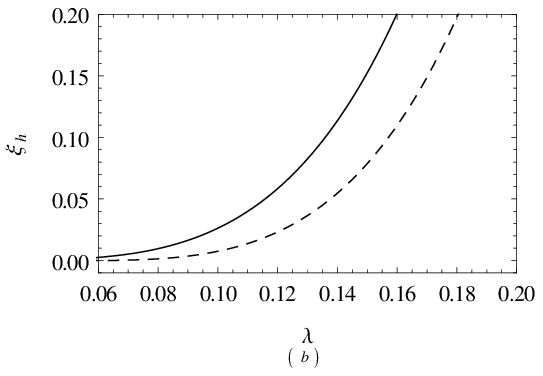}
\end{minipage}
\caption[]{(a) $m_h$ varies with $\lambda$, the solid line and dash-dot line denote $m_h$ and $m_{H_1}$ as $\tan\beta=20$, the dash line and dot line denote $m_h$ and $m_{H_1}$ as $\tan\beta=6$. (b) $\xi_h$ varies with $\lambda$, the solid line and dash line represent as $\tan\beta=20$ and  $\tan\beta=6$, respectively. When $\kappa=0.4$, $A_\lambda=500\;{\rm GeV}$ and $\upsilon_{\nu^c}=2\;{\rm TeV}$.}
\label{fig1}
\end{figure}

When $\kappa=0.4$, $A_\lambda=500\;{\rm GeV}$, and $\upsilon_{\nu^c}=2\;{\rm TeV}$, we plot the lightest Higgs boson mass $m_h$, varying with the parameter $\lambda$ in Fig.~\ref{fig1}(a), where the solid line and dash-dot line denote $m_h$ and $m_{H_1}$ as $\tan\beta=20$, the dash line and dot line denote $m_h$ and $m_{H_1}$ as $\tan\beta=6$, respectively. The mass $m_{H_1}$ denotes the lightest Higgs boson mass if we do not consider the mixing of Higgs doublets and right-handed sneutrinos, and the mass $m_h$ is exactly the lightest Higgs boson mass considering the mixing. The numerical results indicate that the mixing could have significant effects on the lightest Higgs boson mass, as the parameter $\lambda$ is large. With an increase of $\lambda$, the lightest Higgs boson mass $m_h$ drops down quickly, which deviates from the mass $m_{H_1}$. For large $\tan\beta$, the lightest Higgs boson mass $m_h$ decreases more quickly with increasing $\lambda$.

To see the reason more clearly, we also plot the quantity $\xi_h$, varying with $\lambda$ in Fig.~\ref{fig1}(b), where the solid line and dash line, respectively, represent as $\tan\beta=20$ and  $\tan\beta=6$. The quantity $\xi_h$ is defined in Eq.~(\ref{xih}) to quantify the effect on the lightest Higgs boson mass from the mixing of Higgs doublets and right-handed sneutrinos. The figure shows that $\xi_h$ increases quickly with an increase of $\lambda$, and $\xi_h$ for large $\tan\beta$ is larger than it is for small $\tan\beta$. When $\lambda$ is small, $\xi_h$ is also small, and then $m_h$ is close to $m_{H_1}$ because $A_{X_1}^2$ in Eq.~(\ref{AX1}) is in proportion to the parameter $\lambda$. Additionally, in this parameter space, $m_H \approx m_A \approx 2.2\;{\rm TeV}$, $m_{S_R}\approx 1.5\;{\rm TeV}$, and $m_{P_R}\approx 1.1\;{\rm TeV}$, for $\tan\beta=6$ and $\lambda=0.1$. Therefore, for $m_A \sim\mathcal{O}({\rm TeV})$, we can believe that the parameter space is in the large $m_A$ limit, and accordingly the approximate expressions Eq.~(\ref{mH1-app}) and  Eq.~(\ref{mh-app}) will work well. Meanwhile $m_{S_R} \sim\mathcal{O}({\rm TeV})$, and the approximate expression Eq.~(\ref{mh-app1}) is also consistent with the exact one.

\begin{figure}
\setlength{\unitlength}{1mm}
\centering
\begin{minipage}[c]{0.45\textwidth}
\includegraphics[width=2.9in]{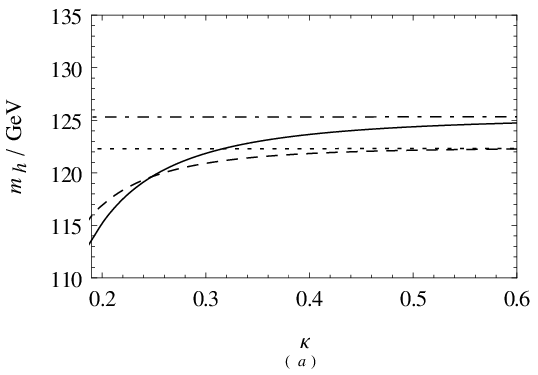}
\end{minipage}%
\begin{minipage}[c]{0.45\textwidth}
\includegraphics[width=2.9in]{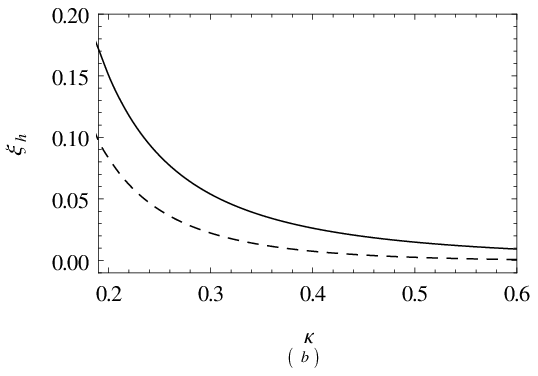}
\end{minipage}
\caption[]{(a) $m_h$ varies with $\kappa$, the solid line and dash-dot line denote $m_h$ and $m_{H_1}$ as $\tan\beta=20$, the dash line and dot line denote $m_h$ and $m_{H_1}$ as $\tan\beta=6$. (b) $\xi_h$ varies with $\kappa$, the solid line and dash line represent as $\tan\beta=20$ and  $\tan\beta=6$, respectively. When $\lambda=0.1$, $A_\lambda=500\;{\rm GeV}$ and $\upsilon_{\nu^c}=2\;{\rm TeV}$.}
\label{fig2}
\end{figure}

We also picture the lightest Higgs boson mass $m_h$ varying with the parameter $\kappa$ in Fig.~\ref{fig2}(a), where the solid line and dash-dot line denote $m_h$ and $m_{H_1}$ as $\tan\beta=20$, the dash line and dot line denote $m_h$ and $m_{H_1}$ as $\tan\beta=6$. And the quantity $\xi_h$ varies with the parameter $\kappa$ in Fig.~\ref{fig2}(b), where the solid line and dash line represent as $\tan\beta=20$ and $\tan\beta=6$, respectively. Here, we take $\lambda=0.1$, $A_\lambda=500\;{\rm GeV}$ and $\upsilon_{\nu^c}=2\;{\rm TeV}$. We can see that the lightest Higgs boson mass $m_h$ deviates from the mass $m_{H_1}$ largely, when the parameter $\kappa$ is small. Of course, for small $\kappa$, the quantity $\xi_h$ is large. Constrained by the Landau pole condition~\cite{mnSSM1}, we choose the parameter $\kappa\leq0.6$.

\begin{figure}
\setlength{\unitlength}{1mm}
\centering
\begin{minipage}[c]{0.45\textwidth}
\includegraphics[width=2.9in]{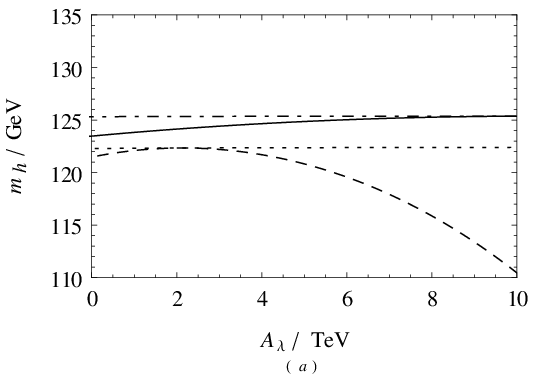}
\end{minipage}%
\begin{minipage}[c]{0.45\textwidth}
\includegraphics[width=2.9in]{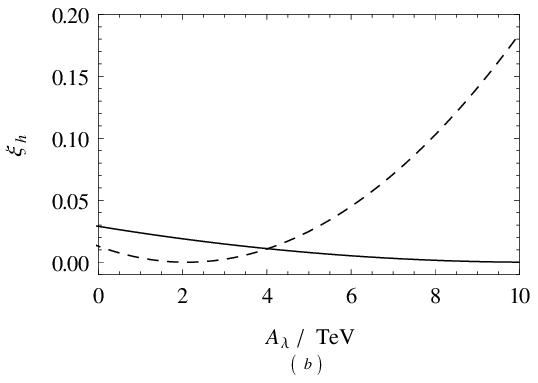}
\end{minipage}
\caption[]{(a) $m_h$ varies with $A_\lambda$, the solid line and dash-dot line denote $m_h$ and $m_{H_1}$ as $\tan\beta=20$, the dash line and dot line denote $m_h$ and $m_{H_1}$ as $\tan\beta=6$. (b) $\xi_h$ varies with $A_\lambda$, the solid line and dash line represent as $\tan\beta=20$ and  $\tan\beta=6$, respectively. When $\kappa=0.4$, $\lambda=0.1$ and $\upsilon_{\nu^c}=2\;{\rm TeV}$.}
\label{fig3}
\end{figure}

In Fig.~\ref{fig3}(a), for $\kappa=0.4$, $\lambda=0.1$ and $\upsilon_{\nu^c}=2\;{\rm TeV}$, we draw the lightest Higgs boson mass $m_h$, varying with the parameter $A_\lambda$, where the solid line and dash-dot line denote $m_h$ and $m_{H_1}$ as $\tan\beta=20$, the dash line and dot line denote $m_h$ and $m_{H_1}$ as $\tan\beta=6$. And Fig.~\ref{fig3}(b) shows the quantity $\xi_h$ versus $A_\lambda$, where the solid line and dash line represent as $\tan\beta=20$ and $\tan\beta=6$, respectively. The numerical results show that $m_h\simeq m_{H_1}$ and $\xi_h\simeq 0$ as $A_\lambda\approx 2\;{\rm TeV}$ for $\tan\beta=6$, and as $A_\lambda\approx 10\;{\rm TeV}$ for $\tan\beta=20$, which is in accordance with Eq.~(\ref{Alambda}). Comparing with the large tree-level contributions, the small one-loop contributions can be ignored, then Eq.~(\ref{Alambda}) can be approximated as
\begin{eqnarray}
A_\lambda \simeq 2\upsilon_{\nu^c} \Big( \frac{3\lambda}{\sin2\beta}  - \kappa \Big) \,.
\label{Alambda1}
\end{eqnarray}
Therefore, when $A_\lambda$ is around $ 2\upsilon_{\nu^c} \Big( {3\lambda}/{\sin2\beta}  - \kappa \Big)$, we could regard the lightest Higgs boson mass as $m_h \approx m_{H_1}$. If $A_\lambda$ drifts off the value of $ 2\upsilon_{\nu^c} \Big( {3\lambda}/{\sin2\beta}  - \kappa \Big)$ significantly, the lightest Higgs boson mass $m_h$ will deviate from the mass $m_{H_1}$.

\begin{figure}
\setlength{\unitlength}{1mm}
\centering
\begin{minipage}[c]{0.45\textwidth}
\includegraphics[width=2.9in]{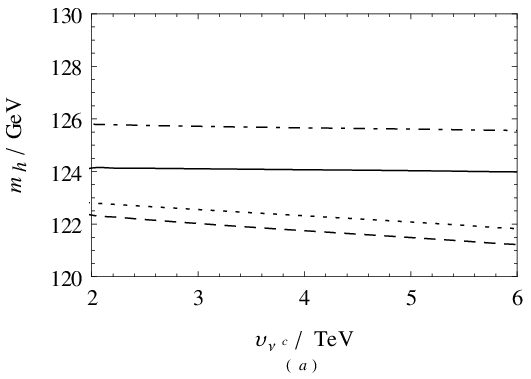}
\end{minipage}%
\begin{minipage}[c]{0.45\textwidth}
\includegraphics[width=2.9in]{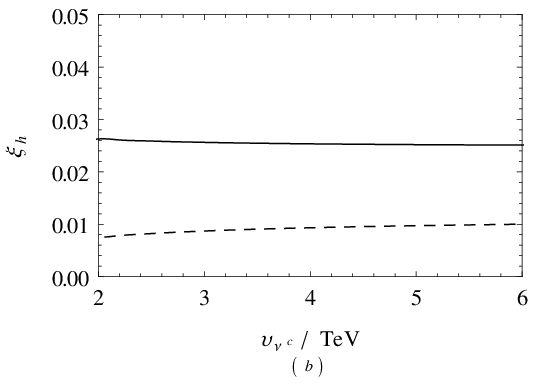}
\end{minipage}
\caption[]{(a) $m_h$ varies with $\upsilon_{\nu^c}$, the solid line and dash-dot line denote $m_h$ and $m_{H_1}$ as $\tan\beta=20$, the dash line and dot line denote $m_h$ and $m_{H_1}$ as $\tan\beta=6$. (b) $\xi_h$ varies with $\upsilon_{\nu^c}$, the solid line and dash line represent as $\tan\beta=20$ and  $\tan\beta=6$, respectively. When $\kappa=0.4$, $\lambda=0.1$ and $A_\lambda=500\;{\rm GeV}$.}
\label{fig4}
\end{figure}

Finally, for $\kappa=0.4$, $\lambda=0.1$, and $A_\lambda=500\;{\rm GeV}$, we plot the lightest Higgs boson mass $m_h$ versus the parameter $\upsilon_{\nu^c}$ in Fig.~\ref{fig4}(a), where the solid line and dash-dot line denote $m_h$ and $m_{H_1}$ as $\tan\beta=20$, and the dash line and dot line denote $m_h$ and $m_{H_1}$ as $\tan\beta=6$. Fig.~\ref{fig4}(b) shows $\xi_h$ varying with $\upsilon_{\nu^c}$, where the solid line and dash line represent as $\tan\beta=20$ and $\tan\beta=6$, respectively. We can see that the lightest Higgs boson mass $m_h$ is parallel to the mass $m_{H_1}$ with increasing of $\upsilon_{\nu^c}$. Through Eq.~(\ref{mR1}), $m_{R_1}^2 \sim \mathcal{O}(\upsilon_{\nu^c}^2)$, and $A_{X_1}^2 \sim \mathcal{O}(\upsilon_{\nu^c})$ as shown in Eq.~(\ref{AX1}). Therefore, the quantity $\xi_h = \frac{(A_{X_1}^2)^2}{m_{R_1}^2 m_{H_1}^2}$ defined in Eq.~(\ref{xih}) becomes flat with an increase of $\upsilon_{\nu^c}$, which can be seen in Fig.~\ref{fig4}(b). In addition, Fig.~\ref{fig4}(a) indicates that $m_h$ and $m_{H_1}$ are decreasing slowly, with an increase of $\upsilon_{\nu^c}$, because the parameter $\mu=3\lambda\upsilon_{\nu^c}$, which can affect the radiative corrections for the lightest Higgs boson mass.

\section{Summary\label{sec-sum}}
\indent
In the framework of the $\mu\nu$SSM, the three singlet right-handed neutrino superfields $\hat{\nu}_i^c$ are introduced to solve the $\mu$ problem of the MSSM. Correspondingly, the right-handed sneutrino VEVs lead to the mixing of the neutral components of the Higgs doublets with the sneutrinos, which produce a large CP-even neutral scalar mass matrix. Therefore, the mixing would affect the lightest Higgs boson mass. In this work, we consider the Higgs boson mass radiative corrections with effective potential methods and then analytically diagonalize the CP-even neutral scalar mass matrix. Meanwhile, in the large $m_A$ limit, we give an approximate expression for the lightest Higgs boson mass seen in Eq.~(\ref{mh-app}). In numerical analysis, we analyze how the key parameters $\lambda,\,\tan\beta,\,\kappa,\,A_\lambda$, and $\upsilon_{\nu^c}$ affect the lightest Higgs boson mass.

\begin{acknowledgments}
\indent
The work has been supported by the National Natural Science Foundation of China (NNSFC)
with Grants No. 11535002, No. 11605037 and No. 11647120,
Natural Science Foundation of Hebei province with Grants No. A2016201010 and No. A2016201069,
Foundation of Department of Education of Liaoning province with Grant No. 2016TSPY10,
Youth Foundation of the University of Science and Technology Liaoning with Grant No. 2016QN11,
Hebei Key Lab of Optic-Eletronic Information and Materials,
and the Midwest Universities Comprehensive Strength Promotion project.
\end{acknowledgments}

\appendix
\section{The masses for the third fermions and their superpartners\label{app1}}
\indent
The masses for the third fermions ${f}={t},{b},{\tau}$ are
\begin{eqnarray}
m_t=Y_t |H_u^0|,\quad  m_b=Y_b |H_d^0|,\quad  m_\tau=Y_\tau |H_d^0|.
\end{eqnarray}
The corresponding $2\times2$ $\tilde{f}_L-\tilde{f}_R$ $(\tilde{f}=\tilde{t},\tilde{b},\tilde{\tau})$ mass squared matrices are
\begin{eqnarray}
M_{\tilde{f}}^2=\left(\begin{array}{ll}M_{\tilde{f}_L}^2&M_{X_f}^2\\
M_{X_f}^{2*}&M_{\tilde{f}_R}^2\end{array}\right),\quad(\tilde{f}=\tilde{t},\tilde{b},\tilde{\tau})
\end{eqnarray}
where the concrete expressions for matrix elements can be given as
\begin{eqnarray}
&&M_{\tilde{t}_L}^2 = m_{\tilde{Q}_3}^2 + \frac{3g_2^2-g_1^2}{12}(|H_d^0|^2-|H_u^0|^2) + Y_t^2 |H_u^0|^2 ,  \\
&&M_{\tilde{t}_R}^2 = m_{\tilde{U}_3}^2 + \frac{g_1^2}{3}(|H_d^0|^2-|H_u^0|^2) + Y_t^2 |H_u^0|^2 ,  \\
&&M_{X_t}^2 =  Y_t (A_t |H_u^0| - \lambda_i \tilde\nu_i^{c*} H_d^0 ) ,  \\
&&M_{\tilde{b}_L}^2 = m_{\tilde{Q}_3}^2 - \frac{3g_2^2+g_1^2}{12}(|H_d^0|^2-|H_u^0|^2) + Y_b^2 |H_d^0|^2 , \\
&&M_{\tilde{b}_R}^2 = m_{\tilde{D}_3}^2 - \frac{g_1^2}{6}(|H_d^0|^2-|H_u^0|^2) + Y_b^2 |H_d^0|^2 ,  \\
&&M_{X_b}^2 = Y_b (A_b H_d^0 - \lambda_i \tilde\nu_i^{c*} H_u^{0*} ),\\
&&M_{\tilde{\tau}_L}^2 = m_{\tilde{L}_3}^2 + \frac{g_1^2-g_2^2}{4}(|H_d^0|^2-|H_u^0|^2) + Y_\tau^2 |H_d^0|^2 , \\
&&M_{\tilde{\tau}_R}^2 = m_{\tilde{E}_3}^2 - \frac{g_1^2}{2}(|H_d^0|^2-|H_u^0|^2) + Y_\tau^2 |H_d^0|^2 ,  \\
&&M_{X_\tau}^2 = Y_\tau (A_\tau H_d^0 - \lambda_i \tilde\nu_i^{c*} H_u^{0*} ).
\end{eqnarray}
Here we ignore the small terms including $Y_{\nu}$ or $|\tilde{\nu_i}|$. The eigenvalues $m_{\tilde{f}_{1,2}}^2$ of the $\tilde{f}=\tilde{t},\tilde{b},\tilde{\tau}$ mass squared matrices can be given by
\begin{eqnarray}
m_{\tilde{f}_{1,2}}^2={{M_{\tilde{f}_L}^2+M_{\tilde{f}_R}^2}\over 2} \pm \sqrt{\Big({{M_{\tilde{f}_L}^2-M_{\tilde{f}_R}^2}\over 2} \Big)^2+|M_{X_f}^2|^2}.
\end{eqnarray}
If substituting the VEVs for the corresponding neutral scalars, the masses of the third fermions ${f}={t},{b},{\tau}$ and their superpartners are manifestly obtained.

\section{The corrections to the minimization conditions\label{app2}}
\indent
Considering one-loop corrections to the minimization conditions from the third fermions ${f}={t},{b},{\tau}$ and their superpartners, $\Delta T_{H_d}$, $\Delta T_{H_u}$, and $\Delta T_{\tilde \nu_{ij}^c} \upsilon_{\nu_j^c}$ are given below:
\begin{eqnarray}
&&\Delta T_{H_d}={3\over(4\pi)^2}\Big\{{G^2\over8}\Big[
f(m_{{{\tilde t}_1}}^2)+f(m_{{{\tilde t}_2}}^2)\Big]
-\Big[Y_{t}^2\mu(A_{t}\tan\beta-\mu)
\nonumber\\
&&\hspace{1.5cm}
-{3g_{2}^2-5g_{1}^2\over24}(m_{{\tilde{t}_L}}^2-m_{{\tilde{t}_R}}^2)\Big]
{f(m_{{{\tilde t}_1}}^2)-f(m_{{{\tilde t}_2}}^2)\over
m_{{{\tilde t}_1}}^2-m_{{{\tilde t}_2}}^2}
\nonumber\\
&&\hspace{1.5cm}
+\Big(Y_{b}^2-{G^2\over8}\Big)\Big[f(m_{{{\tilde b}_1}}^2)+f(m_{{{\tilde b}_2}}^2)\Big]
-2Y_{b}^2f(m_{b}^2)
\nonumber\\
&&\hspace{1.5cm}
+\Big[Y_{b}^2A_{_b}(A_{b}-\mu\tan\beta)
-{3g_{2}^2-g_{1}^2\over24}(m_{{\tilde{b}_L}}^2-m_{{\tilde{b}_R}}^2)\Big]
{f(m_{{{\tilde b}_1}}^2)-f(m_{{{\tilde b}_2}}^2)\over
m_{{{\tilde b}_1}}^2-m_{{{\tilde b}_2}}^2}\Big\}
\nonumber\\
&&\hspace{1.5cm}
+{1\over(4\pi)^2}\Big\{\Big(Y_{\tau}^2-{G^2\over8}\Big)
\Big[f(m_{{{\tilde\tau}_1}}^2)+f(m_{{{\tilde\tau}_2}}^2)\Big]-2Y_{\tau}^2f(m_{\tau}^2)
\nonumber\\
&&\hspace{1.5cm}
+\Big[Y_{\tau}^2A_{\tau}(A_{\tau}-\mu\tan\beta)
-{g_{2}^2-3g_{1}^2\over8}(m_{{\tilde{\tau}_L}}^2-m_{{\tilde{\tau}_R}}^2)\Big]
{f(m_{{{\tilde\tau}_1}}^2)-f(m_{{{\tilde\tau}_2}}^2)\over
m_{{{\tilde\tau}_1}}^2-m_{{{\tilde\tau}_2}}^2}\Big\}
\;,\\
&&\Delta T_{H_u}={3\over(4\pi)^2}\Big\{\Big(Y_{t}^2-{G^2\over8}\Big)
\Big[f(m_{{{\tilde t}_1}}^2)+f(m_{{{\tilde t}_2}}^2)\Big]
-2Y_{t}^2f(m_{t}^2)
\nonumber\\
&&\hspace{1.5cm}
+\Big[Y_{t}^2A_{t}(A_{t}-\mu \cot\beta)
-{3g_{2}^2-5g_{1}^2\over24}(m_{{\tilde{t}_L}}^2-m_{{\tilde{t}_R}}^2)\Big]
{f(m_{{{\tilde t}_1}}^2)-f(m_{{{\tilde t}_2}}^2)\over
m_{{{\tilde t}_1}}^2-m_{{{\tilde t}_2}}^2}
\nonumber\\
&&\hspace{1.5cm}
+{G^2\over8}\Big[f(m_{{{\tilde b}_1}}^2)+f(m_{{{\tilde b}_2}}^2)\Big]
-\Big[Y_{b}^2\mu(A_{b}\cot\beta-\mu)
\nonumber\\
&&\hspace{1.5cm}
-{3g_{2}^2-g_{1}^2\over24}(m_{{\tilde{b}_L}}^2-m_{{\tilde{b}_R}}^2)\Big]
{f(m_{{{\tilde b}_1}}^2)-f(m_{{{\tilde b}_2}}^2)\over
m_{{{\tilde b}_1}}^2-m_{{{\tilde b}_2}}^2}\Big\}
\nonumber\\
&&\hspace{1.5cm}
+{1\over(4\pi)^2}\Big\{{G^2\over8}\Big[f(m_{{{\tilde\tau}_1}}^2)+f(m_{{{\tilde\tau}_2}}^2)\Big]
-\Big[Y_{\tau}^2\mu(A_{\tau} \cot\beta-\mu)
\nonumber\\
&&\hspace{1.5cm}
-{g_{2}^2-3g_{1}^2\over8}(m_{{\tilde{\tau}_L}}^2-m_{{\tilde{\tau}_R}}^2)\Big]
{f(m_{{{\tilde\tau}_1}}^2)-f(m_{{{\tilde\tau}_2}}^2)\over
m_{{{\tilde\tau}_1}}^2-m_{{{\tilde\tau}_2}}^2}\Big\}
\;,\\
&&\Delta T_{\tilde \nu_{ij}^c} \upsilon_{\nu_j^c}={3\over(4\pi)^2}\Big\{ \lambda_i Y_{t}^2 \upsilon_d^2 ( \lambda_j \upsilon_{\nu_j^c} - A_t \tan\beta) {f(m_{{{\tilde t}_1}}^2)-f(m_{{{\tilde t}_2}}^2)\over
m_{{{\tilde t}_1}}^2-m_{{{\tilde t}_2}}^2}
\nonumber\\
&&\hspace{2.1cm}
+ \lambda_i Y_{b}^2 \upsilon_u^2 ( \lambda_j \upsilon_{\nu_j^c} - A_b \cot\beta) {f(m_{{{\tilde b}_1}}^2)-f(m_{{{\tilde b}_2}}^2)\over
m_{{{\tilde b}_1}}^2-m_{{{\tilde b}_2}}^2}\Big\}\nonumber\\
&&\hspace{2.1cm}
+ {1\over(4\pi)^2}\Big\{ \lambda_i Y_{\tau}^2 \upsilon_u^2 ( \lambda_j \upsilon_{\nu_j^c} - A_\tau \cot\beta) {f(m_{{{\tilde \tau}_1}}^2)-f(m_{{{\tilde \tau}_2}}^2)\over
m_{{{\tilde \tau}_1}}^2-m_{{{\tilde \tau}_2}}^2}\Big\}\;,
\end{eqnarray}
with $\mu=\lambda_i \upsilon_{\nu_i^c}$, $f(m^2)=m^2 (\log \frac{m^2}{Q^2}-1)$.

\section{The diagonalization of the $3\times3$ mass matrix\label{app3}}
\indent
The eigenvalues of the $3\times3$ mass squared matrix ${\cal H}$ are given as~\cite{neu-zhang1,top-down}
\begin{eqnarray}
&&m_1^2={a\over3}-{1\over3}p(\cos\phi+\sqrt{3}\sin\phi),
\\
&&m_2^2={a\over3}-{1\over3}p(\cos\phi-\sqrt{3}\sin\phi),
\\
&&m_3^2={a\over3}+{2\over3}p\cos\phi.
\end{eqnarray}
To formulate the expressions in a concise form, one can define the notations,
\begin{eqnarray}
&&p=\sqrt{a^2-3b},\\
&&\phi={1\over3}\arccos({1\over p^3}(a^3-{9\over2}ab+{27\over2}c)),
\end{eqnarray}
with
\begin{eqnarray}
&&a={\rm Tr}({\cal H}),\\
&&b={\cal H}_{11}{\cal H}_{22}+{\cal H}_{11}{\cal H}_{33}+{\cal H}_{22}{\cal H}_{33}
-{\cal H}_{12}^2-{\cal H}_{13}^2-{\cal H}_{23}^2,\\
&&c={\rm Det}({\cal H}).
\end{eqnarray}
In a general way, $m_1^2\leq m_2^2\leq m_3^2$.
So, one can have two possibilities on the mass spectrum:
\begin{itemize}
\item[] (i) spectrum with $m_h<m_H\leq m_{S_3}$:
\begin{eqnarray}
m_h^2=m_1^2,\quad m_H^2=m_2^2,\quad m_{S_3}^2=m_3^2,
\end{eqnarray}

\item[] (ii) spectrum with $m_h<m_{S_3}<m_H$:
\begin{eqnarray}
m_h^2=m_1^2,\quad m_H^2=m_3^2,\quad m_{S_3}^2=m_2^2.
\end{eqnarray}
\end{itemize}

The normalized eigenvectors for the mass squared matrix ${\cal H}$ are given by
\begin{eqnarray}
&&\left(\begin{array}{c}
U_{X_{11}}\\U_{X_{21}}\\U_{X_{31}}
\end{array}\right)
={1\over\sqrt{|X_1|^2+|Y_1|^2+|Z_1|^2}}
\left(\begin{array}{c}X_1\\Y_1\\Z_1\end{array}\right),
\\
&&\left(\begin{array}{c}
U_{X_{12}}\\U_{X_{22}}\\U_{X_{32}}
\end{array}\right)
={1\over\sqrt{|X_2|^2+|Y_2|^2+|Z_2|^2}}
\left(\begin{array}{c}X_2\\Y_2\\Z_2\end{array}\right),
\\
&&\left(\begin{array}{c}
U_{X_{13}}\\U_{X_{23}}\\U_{X_{33}}
\end{array}\right)
={1\over\sqrt{|X_3|^2+|Y_3|^2+|Z_3|^2}}
\left(\begin{array}{c}X_3\\Y_3\\Z_3\end{array}\right),
\end{eqnarray}
with
\begin{eqnarray}
&&X_1=({\cal H}_{22}-m_h^2)({\cal H}_{33}-m_h^2)-{\cal H}_{23}^2, \\
&&Y_1={\cal H}_{13}{\cal H}_{23}-{\cal H}_{12}({\cal H}_{33}-m_h^2), \\
&&Z_1={\cal H}_{12}{\cal H}_{23}-{\cal H}_{13}({\cal H}_{22}-m_h^2),
\\
&&X_2={\cal H}_{13}{\cal H}_{23}-{\cal H}_{12}\Big({\cal H}_{33}-m_H^2\Big),
\\
&&Y_2=({\cal H}_{11}-m_H^2)({\cal H}_{33}-m_H^2)-{\cal H}_{13}^2,
\\
&&Z_2={\cal H}_{12}{\cal H}_{13}-{\cal H}_{23}\Big({\cal H}_{11}-m_H^2\Big),
\\
&&X_3={\cal H}_{12}{\cal H}_{23}-{\cal H}_{13}\Big({\cal H}_{22}-m_{S_3}^2\Big),
\\
&&Y_3={\cal H}_{12}{\cal H}_{13}-{\cal H}_{23}\Big({\cal H}_{11}-m_{S_3}^2\Big),
\\
&&Z_3=({\cal H}_{11}-m_{S_3}^2)({\cal H}_{22}-m_{S_3}^2)-{\cal H}_{12}^2.
\end{eqnarray}

\end{document}